\documentclass[aps,prb,preprint,amsmath,amssymb,superscriptaddress,floatfix]{revtex4-2}
\usepackage{graphicx,xcolor,hyperref}
\usepackage{braket}
\usepackage[ignoreunlbld,norefs,nocites]{refcheck}
\usepackage{import}
\begin{document}

\title{From Entropy to Compression: Competing Thermodynamic Drivers of Structural Transitions in Transition Metals}
\author{S.\ Azadi}
\affiliation{Department of Physics and Astronomy, University of Manchester, Oxford Road, Manchester M13 9PL, UK}
\affiliation{Department of Physics, Clarendon Laboratory, University of Oxford, Parks Road, Oxford OX1 3PU, UK}
\email{sam.azadi@manchester.ac.uk}
\author{S.\ M.\ Vinko}
\affiliation{Department of Physics, Clarendon Laboratory, University of Oxford, Parks Road, Oxford OX1 3PU, UK}
\author{A.\ Principi}
\affiliation{Department of Physics and Astronomy, University of Manchester, Oxford Road, Manchester M13 9PL, UK}
\author{T.\ D.\ K\"{u}hne}
\affiliation{Center for Advanced Systems Understanding, Untermarkt 20, D-02826 G\"orlitz, Germany}
\affiliation{Helmholtz Zentrum Dresden-Rossendorf, Bautzner Landstra{\ss}e 400, D-01328 Dresden, Germany}
\affiliation{TU Dresden, Institute of Artificial Intelligence, Chair of Computational System Sciences, N\"othnitzer Stra{\ss}e 46 D-01187 Dresden, Germany}
\author{M.\ S.\ Bahramy}
\affiliation{Department of Physics and Astronomy, University of Manchester, Oxford Road, Manchester M13 9PL, UK}
\date{\today}

\begin{abstract}
Solid–solid phase transitions in metals are traditionally driven by changes in density or external pressure. Here we show that, under strong electronic excitation, structural stability is governed by the interplay between electronic effects and compression. Using finite-temperature density functional theory, we construct pressure–temperature phase diagrams for 15 metals spanning hcp-, fcc-, and bcc-ground-state structures. The results reveal a systematic reduction of structural diversity with increasing electronic temperature, with stability increasingly dominated by the fcc structure, while hcp remains a persistent secondary phase and bcc stability is progressively suppressed. At elevated temperatures, fcc is broadly favored, whereas bcc is stabilized primarily by compression, leading to a material-dependent competition across the periodic table. These findings provide a unified framework for understanding structural transformations in electronically excited metals and highlight the importance of considering both electronic excitation and pressure in describing phase stability far from equilibrium.
\end{abstract}

\maketitle
\section{Introduction}
The structural and thermodynamic properties of metals are conventionally described in terms of external variables such as pressure, density, and lattice temperature \cite{Issac,Belonoshko2000,Neumann2001,Dubrovinsky,Vocadlo}. Under strong electronic excitation, however, this picture breaks down: the electronic subsystem can be driven far from equilibrium and transiently decoupled from the lattice. In this regime, electrons occupy a broad range of energy states, leading to significant modifications of bonding, screening, and interatomic forces. These changes can induce structural transformations on ultrafast timescales, providing access to states of matter that lie outside conventional equilibrium phase diagrams. Transition metals, with their partially filled d-bands and pronounced structure in the electronic density of states near the Fermi level, are particularly sensitive to such electronic perturbations and therefore offer an ideal platform to explore these effects.

To investigate this regime systematically, we perform a finite-temperature density functional theory (FT-DFT) study of 15 metals spanning hcp-, fcc-, and bcc-ground-state structures. By constructing Gibbs free-energy phase diagrams over a wide pressure–temperature range, we identify clear trends in how structural stability evolves under electronic excitation. Increasing electronic temperature reduces the diversity of stable phases and drives a redistribution of stability toward simpler crystal structures. In particular, the fcc phase becomes increasingly prevalent at elevated temperatures, while the hcp phase persists as a secondary competitor and bcc stability is progressively confined to higher-pressure conditions. Despite these global trends, the detailed phase behavior remains strongly material dependent, reflecting differences in electronic structure and atomic volume across the periodic table.

This behavior can be understood in terms of the competing roles of electronic excitation and mechanical compression. Electronic excitation modifies the distribution of electrons over available states, alters screening, and changes the effective interatomic interactions, favoring more isotropic metallic bonding and stabilizing close-packed structures. In contrast, compression favors structures with smaller atomic volumes, most notably the bcc phase. Structural stability therefore emerges from a balance between these effects, with temperature and pressure selecting among a reduced set of competing crystal structures. This interplay provides a unified interpretation of the trends observed across the calculated phase diagrams.

Experimentally, such conditions can be realized using ultrafast pump–probe techniques, including femtosecond optical lasers and X-ray free-electron lasers (XFELs), which rapidly increase the electronic temperature while leaving the ionic subsystem comparatively cold \cite{Beaurepaire1996,Recoules,Ernstorfer,Humphries,Kang2025,Sundaram,Evans2014,Harb,Celin2025,Amouretti2025,Lee2021,Johnson,Walls,Sokolowski03,MZMo,Fujimoto}. On femtosecond timescales, electron–electron scattering establishes a hot electronic distribution, whereas energy transfer to the lattice occurs more slowly. This separation allows materials to transiently explore high-temperature electronic states at nearly fixed atomic positions, such that structural stability is governed by the electronic free energy. The resulting transformations constitute non-thermal, electronically driven solid–solid phase transitions \cite{Azadi2024,Azadi2025,AzadiFe25,AzadiPRM,AzadiFeSe}, commonly described within the framework of the two-temperature model \cite{TTM,Alexopoulou2024}.

The theoretical description employed here is based on Mermin’s finite-temperature extension of density functional theory \cite{Mermin,Jones2014,Pittalis2011,Dufty2011,Burke2016}, which provides a self-consistent framework for treating temperature-dependent occupations, screening, and exchange–correlation effects. Within this approach, the electronic free energy is obtained directly, enabling a quantitative description of phase stability under electronic excitation. The results presented here establish electronic excitation as a key factor in controlling structural stability in metals and provide a coherent framework for interpreting phase transitions under nonequilibrium conditions. By linking electronic effects with pressure-driven stabilization, this work offers guidance for ultrafast experiments and suggests new routes for controlling material properties through electronic excitation.

\section{Computational details and methods}
\subsection{theoretical background}
Electronic entropy is treated within a finite-temperature formalism through the Fermi–Dirac occupation of electronic states, yielding the entropy contribution to the Gibbs free energy $G=E-TS+PV$,where $E$,$T$,$P$, and $V$ are ground state electronic energy, temperature, external volumetric pressure,and volume of the system, respectively. The entropy $S$ can be written as: 
\begin{equation}
S(T) = -k_B \int n(E)\left[f(E)\ln f(E) + (1-f(E))\ln(1-f(E))\right] dE,
\end{equation}
where $k_B$, $n(E)$, and $f(E)$ are the Boltzmann constant, the electronic density of states, and the Fermi-Dirac distribution function, respectively. The electronic entropy depends on the temperature through the Fermi-Dirac distribution which in general form is given by 
\begin{equation}
    f(E,T)=[1+\text{exp}[(E-\frac{\mu}{k_BT})]]^{-1}
\end{equation}
 with $\mu$ and $E$ as the chemical potential and ground state electronic energy, respectively. For most metals at low and near-room temperatures, the electronic entropy is well described within the Sommerfeld expansion\cite{Ashcroft}, $S = \gamma T$, where the Sommerfeld coefficient $\gamma$ is proportional to the electronic density of states at the Fermi level, $n(E_F)$. In this regime, the entropy contribution is small and varies linearly with temperature, so its influence on structural, elastic, and thermodynamic properties is typically negligible\cite{Grimvall,Porter}. 
 
At elevated electronic temperatures, the entropy term $-TS$ becomes a dominant contribution to the free energy, enabling entropy-driven changes in phase stability. A consistent theoretical description of these effects requires an explicit finite-temperature electronic free-energy formalism, such as FT-DFT. The Kohn-Sham (KS) equation at electronic temperature T, which is solved self-consistently, is given by
 \begin{equation}
     (\frac{-\nabla^2}{2}+V^{T}(\mathbf{r}))\phi_i^{T}(\mathbf{r})=\varepsilon_i^{T}\phi_i^{T}(\mathbf{r})
 \end{equation}
 where the first term is the kinetic energy, $\varepsilon_i^T$ and $\phi_i^{T}(\mathbf{r})$ are KS eigenvalue and eigenfunction, respectively, and 
 \begin{equation}
     V^{T}(\mathbf{r})=V_{ext}(\mathbf{r})+V_{H}^T(\mathbf{r})+V_{xc}^T(\mathbf{r})
 \end{equation}
 where $V_{ext}$, $V_H$, and $V_{xc}$ are external, Hartree, and exchange-correlation potentials, respectively. The density is given by 
 \begin{equation}
     n(\mathbf{r},T)=\sum_i f_i^T|\phi_i^{T}(\mathbf{r})|
 \end{equation}
  with 
  \begin{equation}
      f_i^T=[1+\text{exp}(\frac{\varepsilon_i^{T}-\mu}{T})]^{-1}
  \end{equation}
where $\mu$ is the chemical potential. It should be noted that, although $f_i^T$ is similar to the the Fermi-Dirac function for non-interacting electrons, the difference is that for KS electrons, $f_i^T$ depends also implicitly on T through the KS eigenvalues $\varepsilon_i^T$. 

FT-DFT has proven to be a powerful practical framework for investigating materials under conditions of thermal electronic excitation, most commonly within the generalized-gradient approximation (GGA) for the exchange–correlation (XC) functional \cite{Jones2014}. Although, the conventional local-density approximation (LDA) becomes less accurate in regimes of high electronic temperature and density, recent developments may improve the accuracy of LDA for large electronic densities and finite temperatures \cite{Azadi23,Azadi23II,Karasiev2014,Filinov2015,Schoof2015}.

\subsection{Phase-diagram calculations}
To construct pressure–temperature $(P\text{-}T)$ phase diagrams, we performed systematic FT-DFT calculations for each crystal structure on a dense two-dimensional $(P, T)$ grid. For each element and each structure (hcp, fcc, and bcc), the Gibbs free energy was evaluated over the range $0 < P < 300~\text{GPa}, \quad 0.1 < T < 4.1~\text{eV}$, using pressure increments of 20 GPa and temperature increments of 0.1 eV, resulting in 496 grid points per structure. At each grid point, calculations were carried out at fixed volume corresponding to the target pressure, and the electronic free energy was obtained self-consistently. At each external pressure (corresponding to the horizontal axis of the P–T phase diagrams), the crystal structures and lattice parameters, including the $c/a$ ratio for hcp phases, were first optimized at a low electronic temperature of 0.002 eV while the symmetry of system was conserved.  The stable phase at each $(P, T)$ point was determined by direct comparison of the free energies of the competing structures. To obtain smooth phase boundaries and improve visualization of the phase diagrams, the discrete free-energy data were interpolated using a regular grid interpolation scheme. This procedure provides a continuous representation of the free-energy landscape while preserving the underlying first-principles data. Phase boundaries were extracted from the interpolated free-energy differences between competing phases. The phase diagrams were constructed from $\sim 7500$ FT-DFT calculations on a dense $P\text{-}T$ grid, ensuring a systematic and high-resolution mapping of structural stability.

\subsection{Electronic structure calculations}
All electronic structure calculations were performed using the Quantum ESPRESSO package (version 7.4) \cite{QE}. The Perdew–Burke–Ernzerhof (PBE) generalized-gradient approximation was employed for the exchange–correlation functional \cite{PBE}. PAW pseudopotentials were used, with a plane-wave kinetic-energy cutoff of 120 Ry and an augmentation-charge cutoff of 1200 Ry. Electronic occupations were treated using the Fermi–Dirac distribution at temperature T. The number of empty bands was increased systematically with temperature to ensure convergence of both the internal energy and entropy contributions. Brillouin-zone integrations were performed using Monkhorst–Pack k-point meshes $24\times24\times24$, chosen to ensure convergence of total energies within a few meV per atom. Convergence tests with respect to k-point sampling, plane-wave cutoff, and number of empty bands were performed to ensure that free-energy differences between competing phases are converged within a few meV per atom.

\section{Results and discussion}
The Gibbs free energy phase diagrams of the HCP-ground-state metals display the richest diversity among the three structural classes. Cd and Zn are largely governed by competition between hcp and fcc phases, with bcc appearing only in limited regions, indicating that increasing electronic temperature primarily stabilizes fcc in these systems. In contrast, Mg is predominantly bcc over much of the pressure–temperature range, demonstrating strong volume-driven stabilization of the bcc phase. Ti exhibits full three-phase competition, with hcp favored at low pressure and temperature, bcc stabilized by compression, and fcc emerging at higher temperatures. Zr shows a similarly strong competition between bcc and fcc, while hcp survives only near low-pressure conditions. These results show that HCP-ground-state metals do not follow a single structural response under electronic excitation; instead, they span the full range from close-packed competition to pressure-driven bcc stabilization.

\begin{figure}
    \centering
    \includegraphics[width=1.0\linewidth]{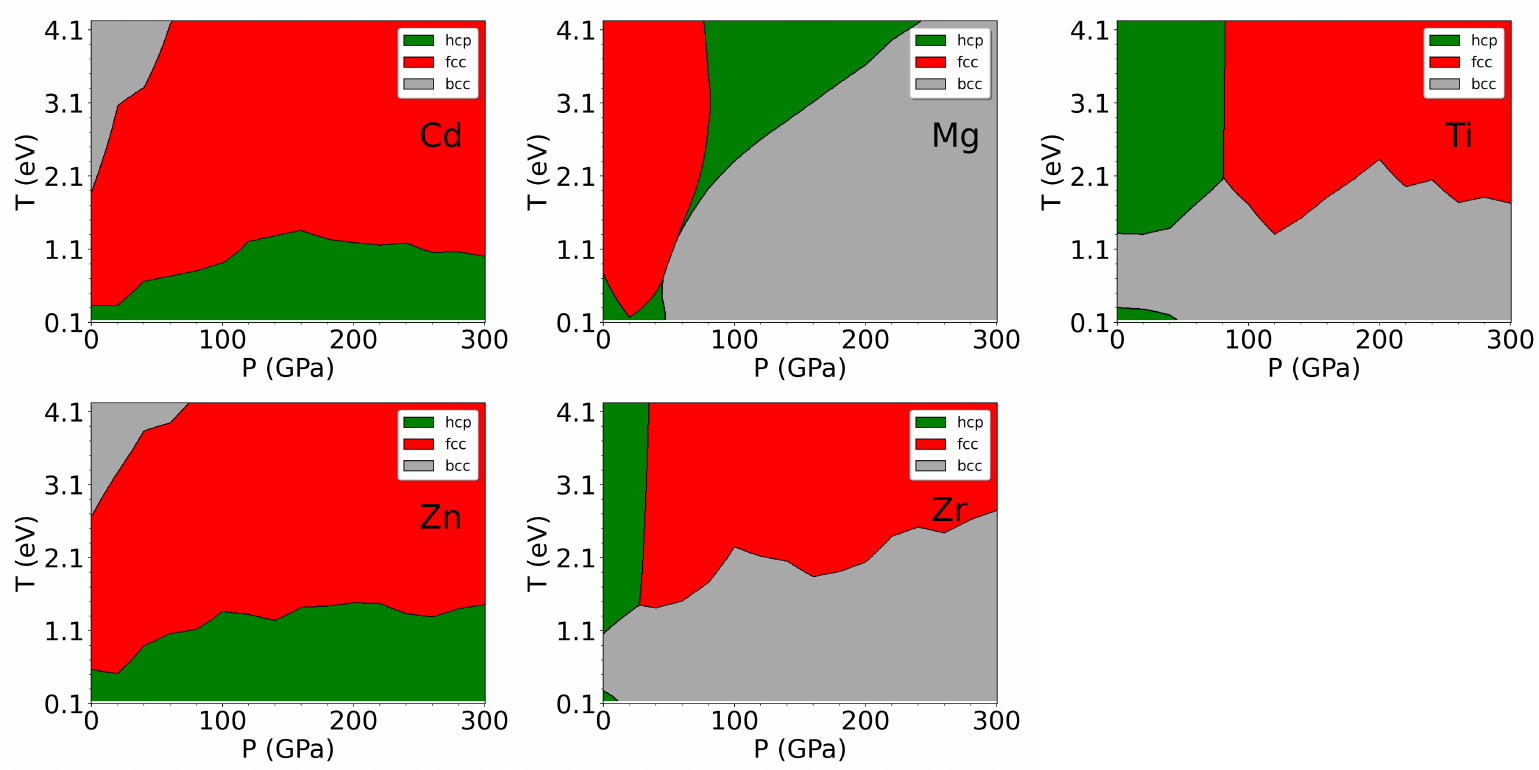}
    \caption{Gibbs free-energy phase diagrams of HCP-ground-state transition metals (Cd, Mg, Ti, Zn, and Zr) as a function of pressure and electronic temperature. At each (P,T) point, the stable phase is determined from the minimum Gibbs free energy among hcp, fcc, and bcc structures. The diagrams reveal a wide range of structural responses under electronic excitation. Cd and Zn exhibit a crossover from hcp at low temperature to fcc over most of the phase space, with only limited bcc stability. In contrast, Mg is largely dominated by the bcc phase, reflecting strong stabilization by compression. Ti displays pronounced three-phase competition, with hcp favored at low pressure, bcc stabilized at higher pressure, and fcc emerging at elevated temperatures. Zr shows a strong pressure-dependent competition between bcc and fcc, while hcp remains stable only near low-pressure conditions.}
    \label{fig:GibsHCP}
\end{figure}

The free energy phase diagrams of the FCC-ground-state metals exhibit a strongly material-dependent behavior. While Al remains a robust fcc system across nearly the entire pressure–temperature range, other elements exhibit substantial competition between alternative structures. Ag, Cu, and Pt are dominated by competition within the close-packed manifold, with pressure promoting the hcp phase over wide regions of the phase diagram, leading to pronounced fcc–hcp coexistence and, in the case of Cu, a broad intermediate-temperature hcp regime. In contrast, Pb displays a qualitatively different behavior, with a large bcc stability region emerging at elevated pressures, indicating a strong influence of volume effects that overcome the preference for close-packed structures. These results demonstrate that phase behavior of FCC-ground-state metals ranges from robust fcc stability to close-packed competition and, in some cases, pressure-driven stabilization of bcc phases.

\begin{figure}
    \centering
    \includegraphics[width=1.0\linewidth]{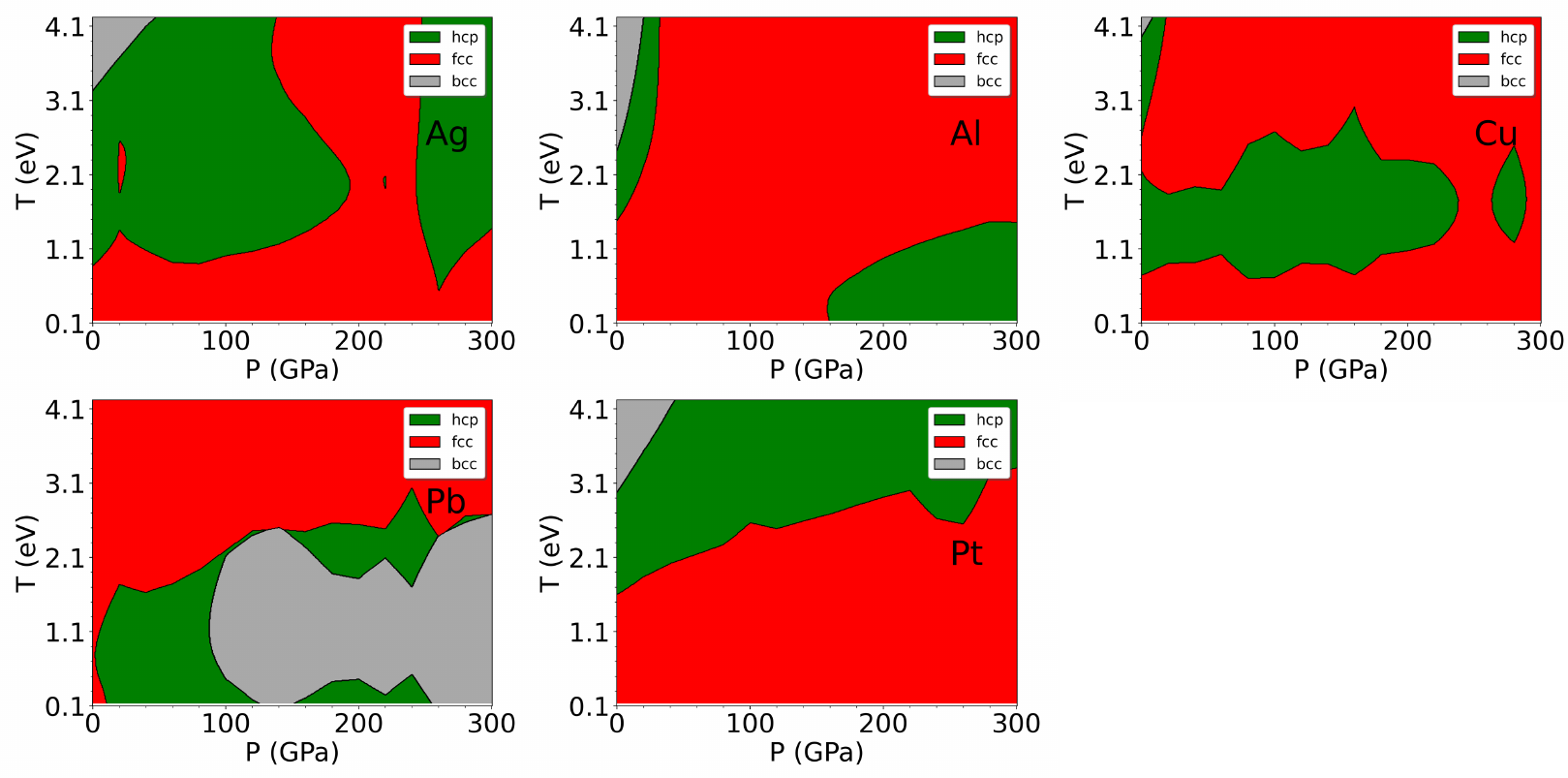}
    \caption{Gibbs free-energy phase diagrams of FCC-ground-state transition metals (Al, Ag, Cu, Pb, and Pt) as a function of pressure and electronic temperature. The stable phase at each (P,T) point is determined from the minimum Gibbs free energy among hcp, fcc, and bcc structures. The diagrams reveal distinct classes of behavior within the FCC group. Al remains robustly fcc across nearly the entire phase space, with only a very limited hcp region at high pressure. Ag, Cu, and Pt exhibit pronounced competition within the close-packed manifold, with pressure promoting the hcp phase over wide regions and, in the case of Cu, a broad intermediate-temperature hcp regime. In contrast, Pb shows a qualitatively different response, with a large bcc stability region emerging at elevated pressures, indicating strong volume-driven stabilization. }
    \label{fig:GibsFCC}
\end{figure}

The phase diagrams of the BCC-ground-state metals show a dominant low-temperature bcc phase in all cases, reflecting the strong stabilization of the lower-volume structure under pressure. With increasing electronic temperature, several systems,including V, Cr, Nb, and W, display a clear expansion of the fcc phase, indicating a temperature-driven crossover toward fcc stability. However, this behavior is not universal: Mo remains predominantly bcc across nearly the entire pressure–temperature range, with only limited hcp stability and negligible fcc participation. These results show that, within the BCC group, structural stability is governed by a competition between pressure-favored bcc and entropy-favored fcc phases, with the strength of this competition varying significantly between materials.

\begin{figure}
    \centering
    \includegraphics[width=1.0\linewidth]{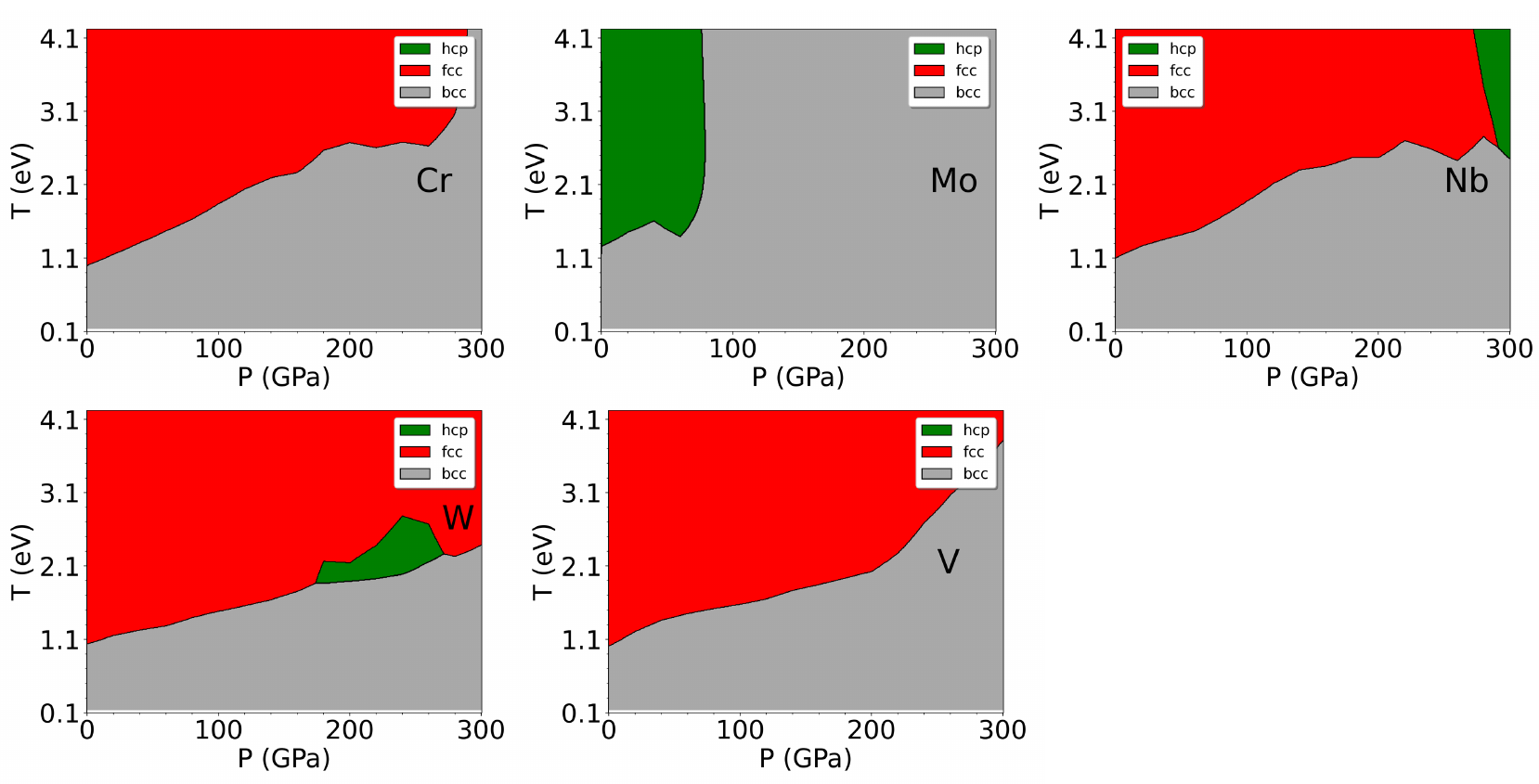}
    \caption{Gibbs free-energy phase diagrams of BCC-ground-state transition metals (Cr, Mo, Nb, W, and V) as a function of pressure and electronic temperature. The stable phase at each (P,T) point is determined from the minimum Gibbs free energy among hcp, fcc, and bcc structures. All systems exhibit a dominant bcc phase at low temperatures, reflecting strong stabilization of the lower-volume structure under compression. With increasing electronic temperature, several metals, including Cr, Nb, W, and V, show a clear expansion of the fcc phase, indicating a temperature-driven crossover toward fcc stability. In contrast, Mo remains predominantly bcc across nearly the entire pressure–temperature range, with only limited hcp stability and negligible fcc participation. The hcp phase appears only in small, localized regions in a subset of systems. }
    \label{fig:GibsBCC}
\end{figure}

The free-energy phase diagrams constructed for the 15 metals provide a detailed and comparative view of how structural stability evolves under electronic excitation. Rather than exhibiting a common response, the diagrams reveal distinct patterns of phase competition that depend on the underlying ground-state structure and atomic volume. In BCC-ground-state metals, the low-temperature bcc phase remains stable at elevated pressures, while increasing electronic temperature promotes the emergence of fcc in several systems, leading to a pressure- and temperature-dependent crossover. FCC-ground-state metals display a wider range of behavior, including cases of robust fcc stability, redistribution of stability within the close-packed manifold (fcc and hcp), and, in elements such as Pb, significant stabilization of bcc under compression. The HCP-ground-state metals exhibit the most diverse phase behavior, with different elements spanning regimes of hcp–fcc competition, strong bcc stabilization, or full three-phase coexistence depending on pressure and temperature. These results show that the phase diagrams provide a consistent framework for identifying how electronic excitation and compression jointly determine structural stability, with the balance between these effects varying significantly across materials.

\begin{figure}
    \centering
    \includegraphics[width=1.\linewidth]{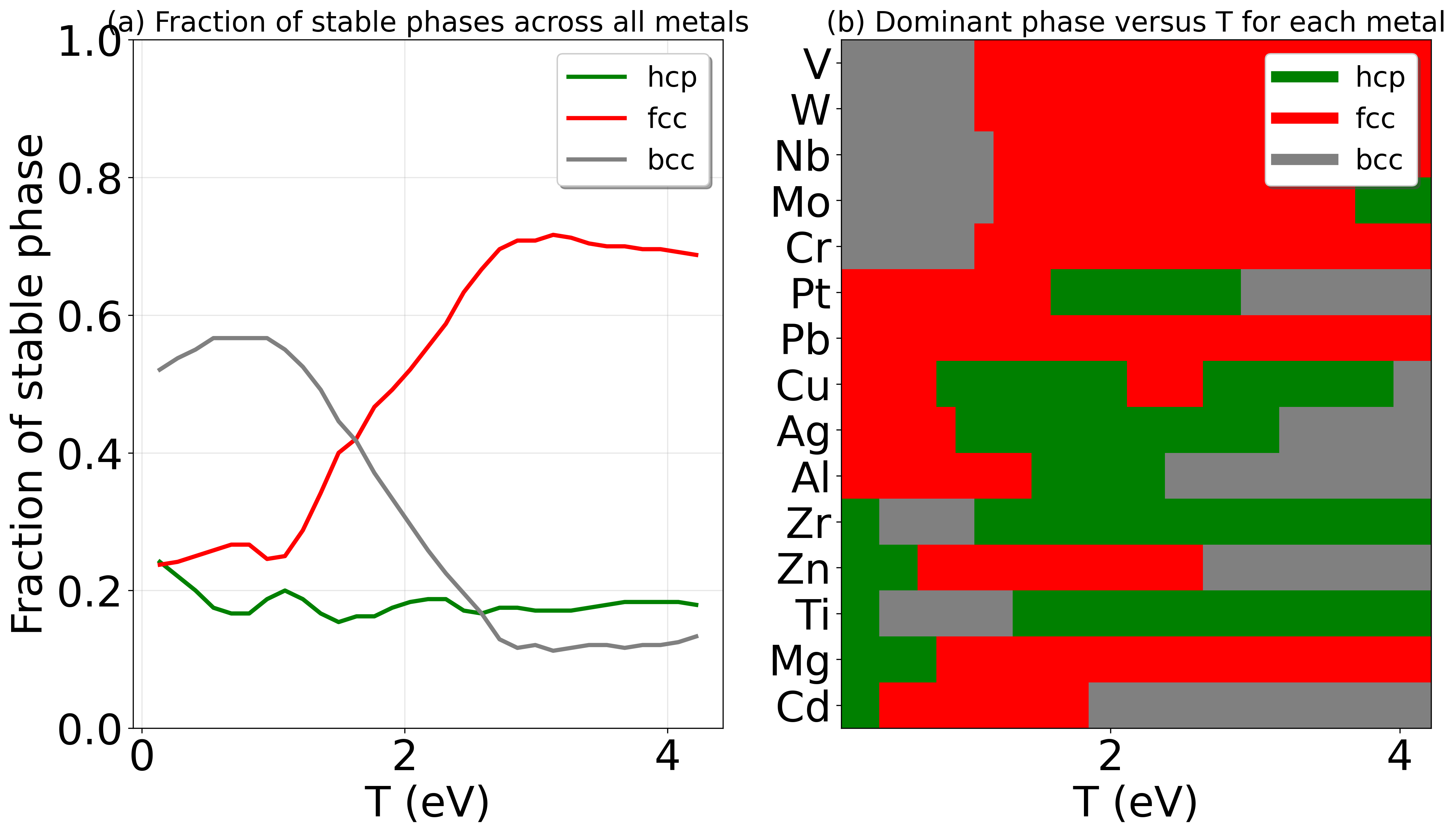}
    \caption{(a) Fraction of stable crystal structures across all 15 metals as a function of electronic temperature, obtained from Gibbs free-energy phase diagrams. At each (P,T) point, the thermodynamically stable phase is identified from the minimum Gibbs free energy among hcp, fcc, and bcc structures, and the fraction is computed over all pressures and elements. Increasing electronic temperature leads to a systematic suppression of bcc stability and a corresponding increase in fcc dominance, while hcp remains a secondary but persistent phase. (b) Dominant crystal structure as a function of electronic temperature for each metal, evaluated at the lowest pressure point of the phase diagram. This representation reflects the temperature-driven structural evolution under near-ambient pressure conditions. Most metals exhibit a transition from their ground-state structure toward either fcc or hcp at elevated temperatures, with fcc emerging as the most prevalent high-temperature phase, while bcc remains stable primarily at low temperatures or in selected materials.}
    \label{fig:gibssummary}
\end{figure}

Figure \ref{fig:gibssummary} summarizes the global trends in phase stability extracted from the free-energy calculations. Panel (a) shows that increasing electronic temperature drives a pronounced redistribution of stability across all metals: the fraction of bcc phases decreases rapidly, while fcc becomes the dominant structure above $T \sim 0.15–0.20$ Ry, reaching $\sim 70\%$ of the phase space at the highest temperatures. The hcp fraction remains relatively constant, indicating that it plays a secondary but persistent role. This behavior reflects a systematic weakening of the pressure-stabilized bcc phase under strong electronic excitation and a concurrent stabilization of close-packed structures, particularly fcc. Panel (b) provides a complementary, physically intuitive view by tracking the dominant phase at low pressure for each metal. At low temperatures, the expected ground-state structures are recovered, confirming the consistency of the approach. As temperature increases, many systems undergo transitions toward fcc or, in some cases, hcp, while bcc stability is progressively restricted to a smaller subset of materials. Together, these panels demonstrate that strong electronic excitation reduces structural diversity and drives a convergence toward a simplified structural landscape dominated by close-packed phases, with fcc emerging as the most prevalent high-temperature structure, while the detailed pathway remains strongly material dependent.

\section{Conclusion and outlook}
We have investigated electronically driven solid–solid phase transitions in transition metals using finite-temperature density functional theory, constructing Gibbs free-energy phase diagrams for 15 elements over a wide pressure–temperature range. The results provide a comprehensive picture of how structural stability evolves under electronic excitation, revealing a progressive reduction in the number of competing phases as temperature increases. This evolution is accompanied by a redistribution of stability toward simpler crystal structures, most notably fcc, while bcc stability becomes increasingly restricted to higher-pressure conditions and hcp persists as a secondary phase in selected systems.

The detailed phase behavior depends strongly on the ground-state structure and atomic volume. BCC-ground-state metals generally retain bcc stability at low temperatures and pressures, with fcc emerging at elevated temperatures in several cases. FCC-ground-state metals display a broader range of responses, including robust fcc stability, competition within the close-packed manifold, and, in some cases, pressure-driven stabilization of bcc. HCP-ground-state metals exhibit the widest diversity, spanning regimes of hcp–fcc competition, strong bcc stabilization, and three-phase coexistence. These results show that structural stability under electronic excitation is determined by the balance between electronic effects and compression, which together select among a reduced set of crystal structures and define a material-dependent phase landscape.

A detailed comparison between phase diagrams constructed from the Helmholtz free energy $F$ and the Gibbs free energy $G$ (see Supplementary Information \cite{Suppl}) provides insight into the relative importance of the entropy term $-T\Delta S$ and the pressure–volume contribution $P\Delta V$ in determining phase stability. In the high-temperature, low-pressure regime, both approaches yield consistent phase boundaries, indicating that stability is primarily governed by entropy-driven effects. In contrast, significant deviations emerge at elevated pressures, where the inclusion of the $P\Delta V$ term alters the relative stability of competing phases, most notably stabilizing bcc structures with smaller atomic volumes. These comparisons show that the trends observed in the Gibbs phase diagrams arise from a balance between entropy-driven stabilization at high temperature and compression-driven effects at high pressure, with the dominant contribution varying across the pressure–temperature space.

More broadly, this work introduces electronic entropy as an additional thermodynamic dimension in phase stability and equations of state. Under strong electronic excitation, entropy can modify structural stability independently of density, shifting phase boundaries relative to conventional compression pathways. Such effects may contribute to discrepancies between static and dynamic compression experiments \cite{Celin2025,Amouretti2025}. Capturing this physics in predictive simulations will require the development of interatomic potentials that explicitly incorporate electronic excitation and entropy. At the same time, emerging ultrafast pump–probe techniques, combining femtosecond optical or X-ray excitation with time-resolved diffraction and spectroscopy, provide a promising route to experimentally probe these entropy-driven structural transformations and to test the universal behavior identified here.

\section{Acknowledgment}
We acknowledge the support of the Leverhulme Trust under the grant agreement RPG-2023-253. S. Azadi and T.D. K\"{u}hne acknowledge the computing time provided to them on the high-performance computers at the NHR Center in Paderborn (PC2). 

\section{Conflicts Of Interest}
The authors declare no conflicts of interest.

\section{Data Availability Statement}
The data that support the findings of this study are available from thecorresponding author upon reasonable request.

\bibliography{main_arxiv}
\begin{center}
    \Huge Supplementary Information
\end{center}
\section{Helmholtz phase diagram}
To elucidate the thermodynamic origin of the phase behavior, we compare phase diagrams constructed from the Helmholtz and Gibbs free energies. The Helmholtz formulation, isolates the contribution of electronic entropy through the $-T\Delta S$ term, allowing us to assess how increasing electronic temperature modifies relative phase stability in the absence of volume relaxation. In contrast, the Gibbs formulation incorporates the additional $P\Delta V$ contribution associated with externally applied pressure, enabling structural selection through differences in atomic volume between competing phases. The comparison between these two descriptions therefore provides a direct means of disentangling the competition between entropy-driven stabilization and pressure-induced volume effects in determining phase stability. It should be emphasized that only Gibbs free energy phase diagrams correspond to constant-pressure conditions and are therefore appropriate for direct comparison with experimental phase boundaries.

We performed the same analysis for all 15 elements using the Helmholtz free energy $F$ in addition to the Gibbs free energy $G$. The relationship between the two thermodynamic potentials is given by
$G(P,T) = F(V,T) + P V$,
where $V$ is the volume per atom. Consequently, the difference in stability between two competing phases $\alpha$ and $\beta$ can be written as
$\Delta G_{\alpha\beta}(P,T) = \Delta F_{\alpha\beta}(T) + P\,\Delta V_{\alpha\beta}$,
with $\Delta F_{\alpha\beta} = F_\alpha - F_\beta$ and $\Delta V_{\alpha\beta} = V_\alpha - V_\beta$. The Helmholtz free-energy difference $\Delta F_{\alpha\beta}$ contains both internal energy and entropy contributions,
$\Delta F_{\alpha\beta}(T) = \Delta E_{\alpha\beta} - T\,\Delta S_{\alpha\beta}$,
so that phase stability is determined by the competition between the entropy term $-T\Delta S$ and the pressure–volume term $P\Delta V$.

Comparing phase diagrams constructed from $F$ and $G$ therefore provides direct insight into the relative importance of these contributions. At a given $(P,T)$ point, when both $F$ and $G$ predict the same stable phase, the $P\Delta V$ term is negligible and phase stability is dominated by the entropy-driven contribution contained in $\Delta F$. In contrast, discrepancies between the two phase diagrams identify regions where $P\Delta V$ is comparable to or larger than $-T\Delta S$, leading to shifts in phase boundaries or changes in the stable structure. This comparison allows us to isolate the role of compression in modifying electronically driven phase stability and to quantify how volume differences between competing phases influence the Gibbs phase diagrams.

\begin{figure}
    \centering
    \includegraphics[width=1.0\linewidth]{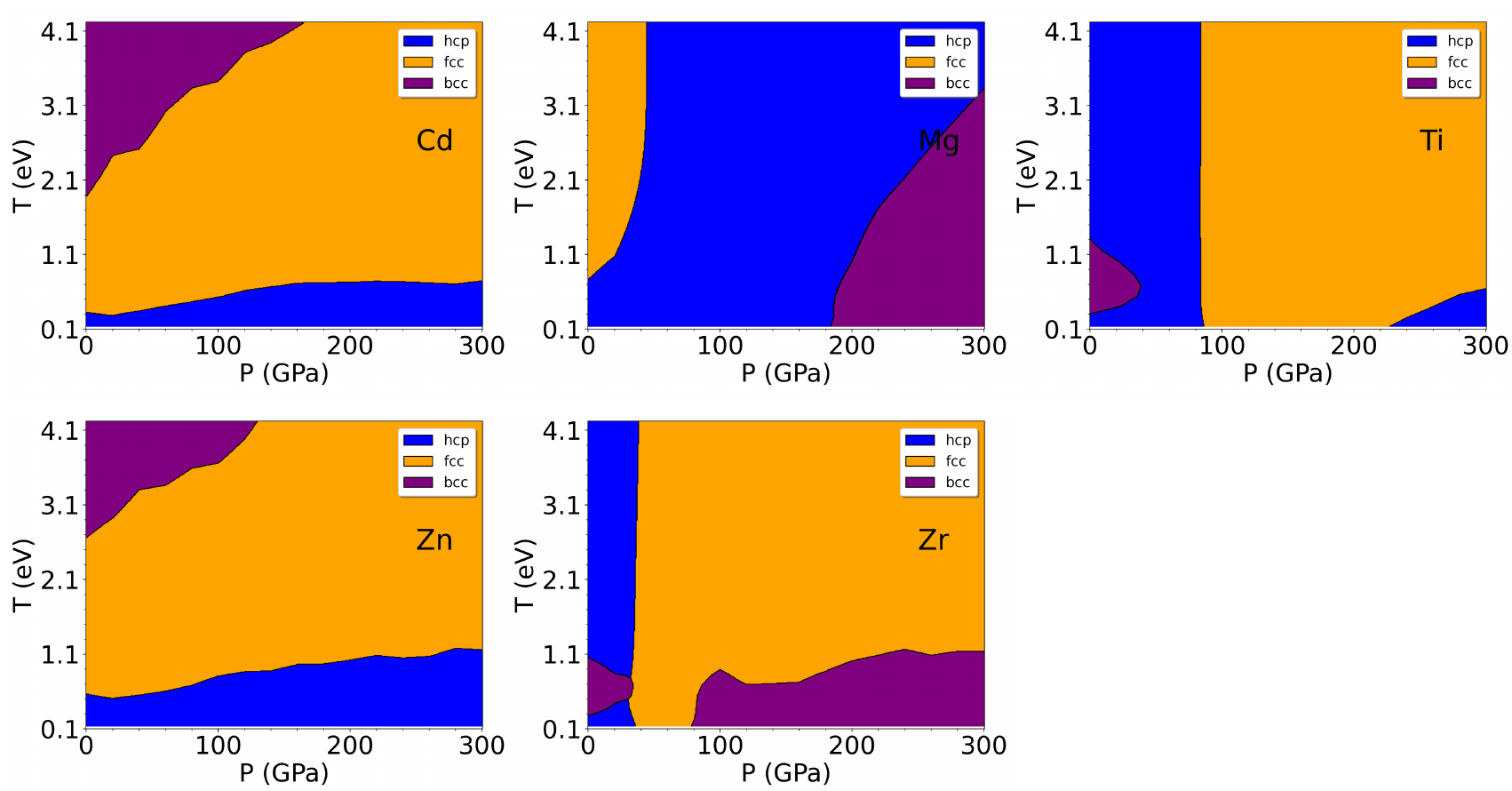}
    \caption{Pressure–temperature Helmholtz phase diagrams of hcp-ground-state metals (Cd, Mg, Ti, Zn, and Zr) obtained from FT-DFT. Increasing electronic temperature drives collapse of the hcp stability field and promotes the emergence of fcc as the dominant structure over wide pressure ranges. Despite the diversity of ground-state behavior, all systems evolve toward a reduced structural manifold characterized by close-packed phases, with Mg representing a notable exception due to its weak d-electron character. }
    \label{fig:HCPPTS}
\end{figure}

For the hcp-ground-state metals (Cd, Mg, Ti, Zn, and Zr), the calculated Helmholtz phase diagrams (Fig.~\ref{fig:HCPPTS}) reveal that the stability of the hcp phase is largely confined to low electronic temperatures and limited pressure ranges. With increasing temperature, most systems exhibit a progressive reduction of the hcp stability field, accompanied by the emergence and eventual dominance of the fcc phase. This trend is particularly pronounced in Ti, Zr, Zn, and Cd, where fcc becomes the energetically preferred structure over a broad region of the phase diagram at elevated temperatures. The primary exception is Mg, which retains a relatively large hcp stability region across the explored pressure range, reflecting its weaker d-electron character and more free-electron-like bonding. The results indicate that electronic excitation tends to destabilize the hcp phase and promote competition within the close-packed structural family.

\begin{figure}
    \centering
    \includegraphics[width=1.0\linewidth]{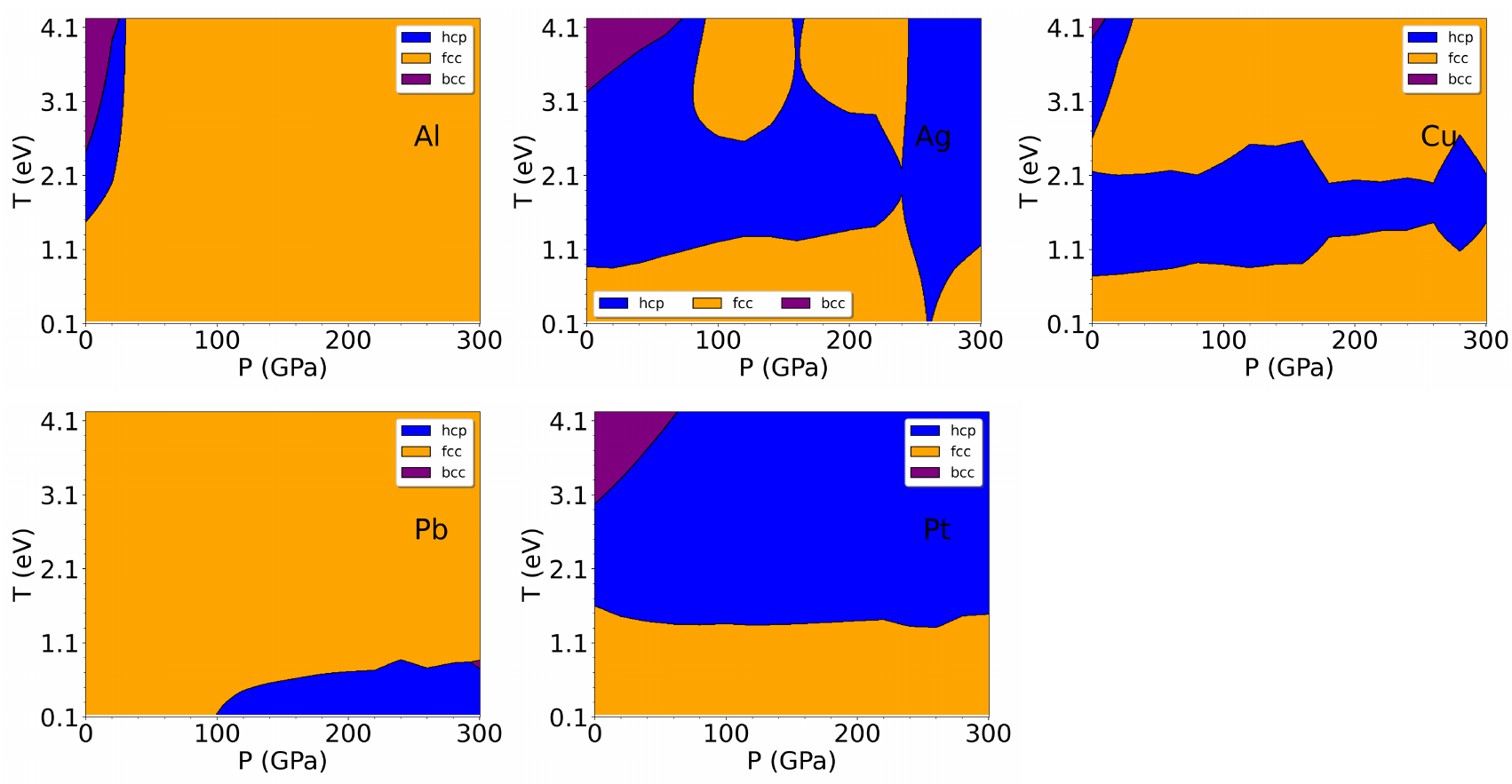}
    \caption{Pressure–temperature Helmholtz phase diagrams of fcc-ground-state metals (Al, Ag, Cu, Pb, and Pt) obtained from FT-DFT. The fcc phase remains remarkably robust across a wide range of pressures and electronic temperatures, with only limited competition from hcp at elevated temperatures and intermediate pressures. Despite strong electronic excitation, no significant stabilization of non-close-packed phases is observed. }
    \label{fig:FCCPTS}
\end{figure}

For the fcc-ground-state metals (Al, Ag, Cu, Pb, and Pt), the phase diagrams (Fig.~\ref{fig:FCCPTS}) show that the fcc structure remains robust across a wide range of pressures and temperatures, with only limited competition from hcp or bcc phases. In particular, Al and Pb exhibit an almost complete dominance of the fcc phase throughout the explored thermodynamic space, indicating that strong electronic excitation does not significantly alter their structural preference. In contrast, Ag, Cu, and Pt display a more pronounced competition between fcc and hcp structures, especially at elevated temperatures and intermediate pressures. Nevertheless, this competition remains confined to the close-packed family, and no substantial stabilization of bcc phases is observed. These results demonstrate that for fcc-ground-state metals, electronic entropy preserves or reinforces close-packed stability rather than inducing qualitatively new structural phases.

\begin{figure}
    \centering
    \includegraphics[width=1.0\linewidth]{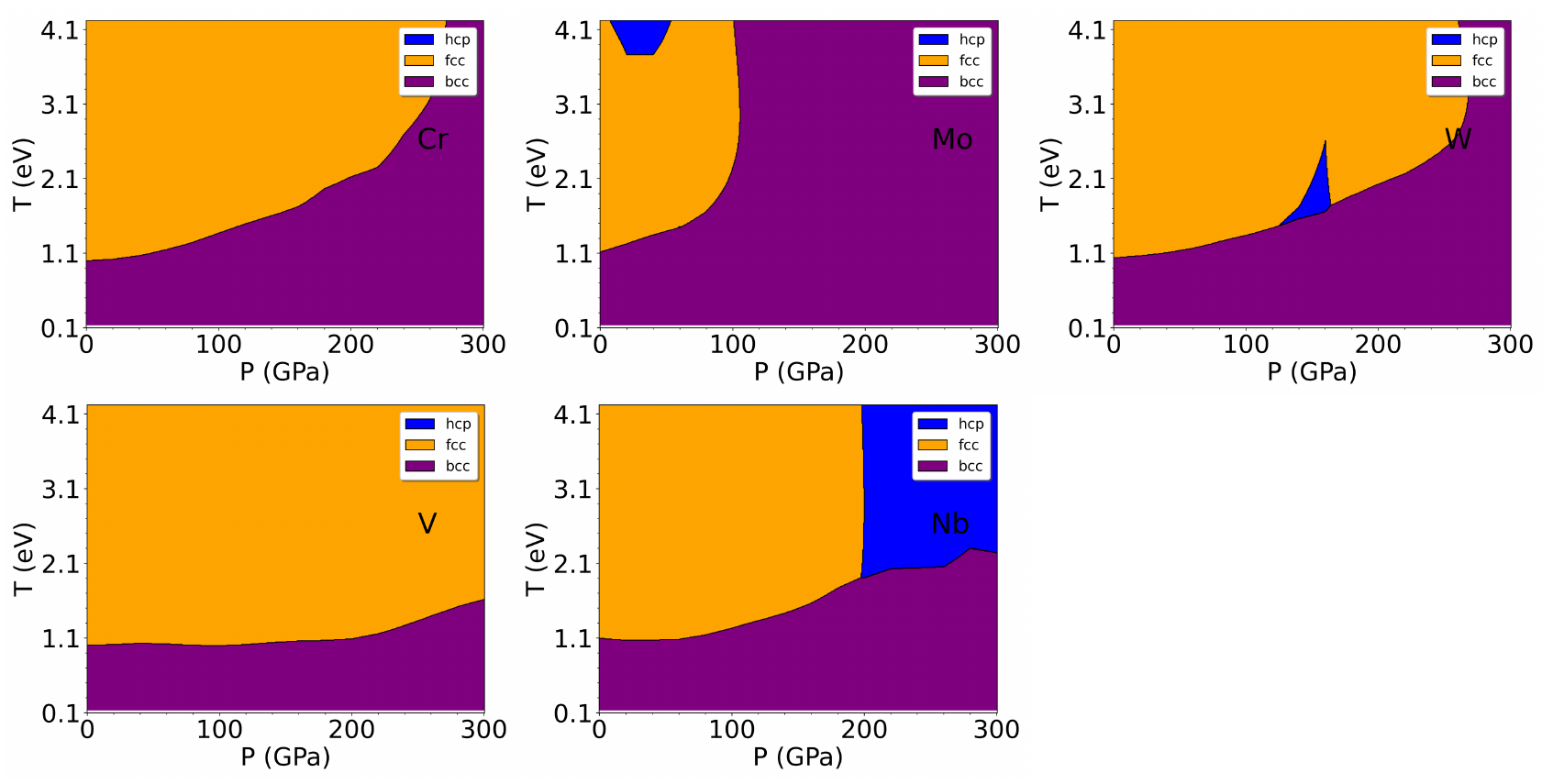}
    \caption{Pressure–temperature Helmholtz phase diagrams of bcc-ground-state metals (Cr, Mo, W, V, and Nb) obtained from FT-DFT. Increasing electronic temperature leads to a pronounced reduction of the bcc stability field and the emergence of competing close-packed phases. At elevated temperatures, fcc and hcp structures occupy an increasingly large portion of the phase space, indicating a strong entropy-driven reorganization of structural stability. Even in refractory systems where bcc remains comparatively robust, the phase space is no longer exclusively bcc-dominated. }
    \label{fig:BCCPTS}
\end{figure}

For the bcc-ground-state metals (Cr, Mo, W, V, and Nb), the phase diagrams, shown in Fig.~\ref{fig:BCCPTS}, exhibit a markedly different low-temperature behavior but converge toward a similar high-temperature regime. At low electronic temperatures, the bcc phase is dominant across a wide pressure range, consistent with the known ground-state structures of these elements. However, with increasing temperature, the stability of the bcc phase is progressively reduced, and regions of fcc and, in some cases, hcp stability emerge and expand. This effect is particularly evident in V and Nb, where close-packed phases occupy a significant portion of the phase diagram at elevated temperatures. Even in refractory systems such as Mo and W, where bcc remains comparatively robust, the phase space is no longer exclusively bcc-dominated. These results indicate that electronic excitation drives a redistribution of structural stability away from the bcc phase toward close-packed configurations.

\begin{figure}
    \centering
    \includegraphics[width=1.0\linewidth]{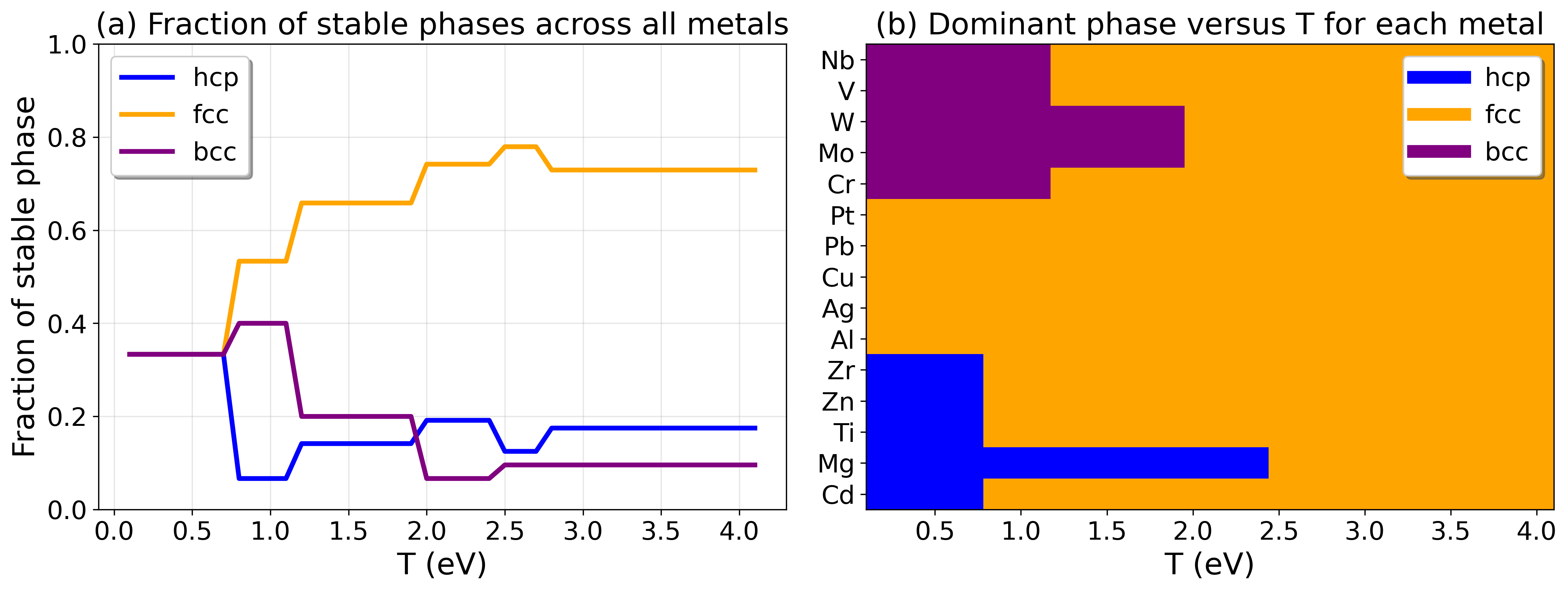}
    \caption{Summary of the structural evolution of 15 metals under strong electronic excitation using Helmholtz free energy data. (a) Fraction of stable phases as a function of electronic temperature, obtained by counting the number of pressure points and elements for which hcp, fcc, or bcc is the thermodynamically stable structure. (b) Dominant phase as a function of electronic temperature for each metal, determined from the majority stable phase over the full pressure range considered. }
    \label{fig:universal}
\end{figure}

Figure~\ref{fig:universal} provides a compact summary of the structural evolution of the 15 metals under increasing electronic temperature. Panel (a) shows the fraction of stable phases obtained by sampling all pressures and elements at each temperature, while panel (b) presents the dominant phase for each metal as a function of temperature. At low temperatures, the distribution of phases reflects the diversity of ground-state structures, with hcp-, fcc-, and bcc-based stability regions clearly distinguished. As the temperature increases, however, both panels reveal a systematic reorganization of phase stability: the prevalence of the bcc phase decreases, while the close-packed phases become increasingly dominant. Consistently, panel (b) shows that metals with initially different ground-state structures progressively converge toward fcc or hcp stability at elevated temperatures.


\end{document}